# Spatial Scales of Living Cells, their Energetic and Informational Capacity and the Origin of "the Arrow of Time"

Edward Bormashenko[1], Alexander Voronel[2]

[1]*Ariel University, Department of Chemical Engineering, Biotechnology and Materials, Engineering Faculty, P.O.B. 3, 407000, Ariel, Israel*

[2]*Tel-Aviv University, Ramat Aviv, 66978 Tel-Aviv, Israel*

Corresponding author:

Edward Bormashenko, Department of Chemical Engineering, Biotechnology and Materials, Engineering Faculty.

Ariel University,

P.O.B. 3

Ariel 40700

Phone: +972-3-906-6134

Fax: +972-3-906-6621

E-mail: edward@ariel.ac.il

**Abstract**

Physical (thermodynamic and kinetic), chemical and biological reasoning restrict the spatial dimensions of living cells and confine them to between one and one hundred micrometers. Cells should necessarily be macroscopic, dissipative objects, resisting thermal fluctuations and providing sufficient informational capacity. The upper limit of spatial dimensions of cells is supplied by their ability to withstand gravity and inertia forces under reasonable deformations. The upper limit of cell dimensions is also governed by the hierarchy of characteristic time scales, inherent for mass and heat transport. For micron-scaled cells, the "traffic time" (namely a typical time necessary for the migration of one enzyme to another) is on the order of magnitude of a millisecond, which coincides with the characteristic time scale of a single round of the catalytic enzyme cycle. Among other important consequences the macroscopic dimensions of living cells (seen as dissipative systems) give rise to the irreversibility of biological processes, demonstrating "the arrow of time".

**Introduction**

The physical behavior of living cells has been subjected to extensive research during the past decade [1-4]. Much has been gained in the understanding of perplexing energetic [5-6] and informational [7-8] exchanges between living cells. As always it happens that scaling arguments become extremely useful for the analysis of complicated phenomena, when accurate solutions of differential equations, related to the bio-system, become unavailable [9-11].

In the presented paper, we address fundamental scaling laws, describing the physical behavior of living cells. We note, that the living cells appearing in the vast majority of biological systems have a characteristic dimension $l$, confined to between 1 and 100 μm [12-13]. Even exceptions such as *Acetabularia mediterranea*, a single-celled organism, which is gigantic in size, possess a cross-section with a characteristic dimension of 100 μm [12]. However, it remains unclear: how a cell's size is determined. Several hypotheses have been proposed to explain the narrow range of characteristic spatial dimensions of living cells, including mechanistic explanations [3] as well as reasoning that focus on the crucial role of water supply for defining the dimensions of living cells [14]. We suggest a physical reasoning based on fundamental scaling laws to explain the micro-scaled dimensions of living cells. Our proposed reasoning allows an estimation of the informational capacity of living cells, and supplies one of the possible explanations for "the arrow of time", which is inherent for biological systems.

## 2. Scaling laws governing the behavior of living cells

*2.1. Interrelation between bulk and surface energies of living cells*

Consider the interrelation between the volume (bulk) $E_V$ and the surface $E_S$ energies of living cells, described by the dimensionless ratio:

$$\varsigma = \frac{E_V}{E_S}. \tag{1}$$

Assuming that $E_V \sim \varepsilon l^3$ and $E_S \sim \gamma_{int} l^3$ (where $\varepsilon$ and $\gamma_{int}$ are the densities of the volume and surface (interfacial) energies of a cell with the dimensions of: $[\varepsilon] = \frac{J}{m^3}$ and $[\gamma_{int}] = \frac{J}{m^2}$ respectively, and $l$ is the characteristic dimension of a cell), yields the following expression for a rough estimation of the dimensionless ratio $\varsigma$:

$$\varsigma \cong \frac{\varepsilon}{\gamma_{int}} l. \tag{2}$$

It is reasonable to assume for the value of $\varepsilon$ the volume density of the hydrogen bonding, calculated as follows:

$$\varepsilon \cong \frac{\varepsilon_{mol} \overline{\rho}}{\overline{\mu}}, \tag{3}$$

where $\varepsilon_{mol}$ is the molar energy of hydrogen bonding, and $\overline{\rho}$ and $\overline{\mu}$ are the averaged density and molecular weight of a cell respectively. Assuming for the sake of a rough estimation that $\varepsilon_{mol} \cong 23 \times 10^3 \frac{J}{mol}$ (see Ref. 15), $\overline{\rho} \cong 1.0 \times 10^3 \frac{kg}{m^3}$ and $\overline{\mu} \cong 18 \times 10^{-3} \frac{kg}{mol}$, we obtain: $\varepsilon \cong 1.3 \times 10^9 \frac{J}{m^3}$. The specific interfacial surface energy of cells (in other words the specific surface energy of a cell membrane) is not a well-established physical value; however, it may be crudely estimated as $\gamma_{int} \cong (1-10) \times 10^{-3} \frac{J}{m^2}$ (see Refs. 16, 17). Substituting the aforementioned values of the physical parameters in Eq. 2, and assuming that $l \cong (1-100) \times 10^{-6} m$, yields for the dimensionless ratio $\varsigma$ the estimation: $1.3\times 10^5 < \varsigma < 1.3\times 10^8$. This means, that in living cells the internal energy of a bulk always prevails over the surface energy. As seen from Eq. 2 large values of $\varsigma$, growing with the characteristic dimensions of cells, provide living cells with their existence as physical entities. However, these dimensions do not grow to infinity, being restricted by external gravity and inertia forces, as demonstrated in the next Section.

*2.2. Gravity (inertia) forces and the spatial dimensions of living cells*

The interrelation between gravity (inertia) forces and interfacial phenomena is supplied by the so-called Bond numbers:

$$Bo_g = \frac{\Delta\rho g l^2}{\gamma_{int}}, \qquad (4a)$$

$$Bo_{in} = \frac{\Delta\rho a l^2}{\gamma_{int}}, \qquad (4b)$$

where $Bo_g$ and $Bo_{in}$ are the gravitational and inertial Bond numbers respectively, $g$ and $a$ are the accelerations of gravity and of a moving body respectively and $\Delta\rho$ is the difference in the densities of the two phases [18-20].

For an estimation of the maximal values of the Bond numbers we anticipate: $\Delta\rho \cong \bar{\rho} \cong 1.0\times 10^3 \frac{kg}{m^3}; g \cong a \cong 10\frac{m}{s^2}; \gamma_{int} \cong 10^{-3}\frac{J}{m^2}; l \cong 10^{-4}m$. Substituting these values into Eqs. 4a-b results in the following estimations of the maximal values of Bond numbers:

$$Bo_g^{max} \cong Bo_{in}^{max} \cong 0.1. \qquad (5)$$

This means that interfacial phenomena are dominant upon external effects due to gravity or inertia. This situation allows cells to move with accelerations $a \le g$ without essential deformations, that would result resulting in disrupting their biological functioning [21-22]. However, for $l \cong 10^{-3}m$ we estimate $Bo_g^{max} \cong Bo_{in}^{max} \cong 10$ and the effects due to gravity/inertia cannot be compensated by interfacial effects (it would result in deformations, which may be deadly for cells). However, large cells (namely $l \ge 10^{-3}m$) may survive in liquids, where the values of $\Delta\rho$ are low, and this is the case for giant single-celled organisms (weeds) such as *Acetabularia mediterranea* [12]. Therefore, the upper values of the spatial dimensions of living cells are restricted by gravity and acceleration, as is seen from Eqs. 4a-b.

*2.3. Spatial dimensions of living cells and thermal fluctuations; an informational capacity of living cells*

Spatial dimensions of living cells being confined to values between 1 and 100 μm, provide validity to Eqs. 5a-b:

$$\frac{E_V}{k_B T} \cong \frac{\varepsilon l^3}{k_B T} \cong 3 \times 10^{11} - 3 \times 10^{17} \gg 1, \quad (5a)$$

$$\frac{E_S}{k_B T} \cong \frac{\gamma_{int} l^2}{k_B T} \cong 2.5 \times 10^5 - 2.5 \times 10^{10} \gg 1, \quad (5b)$$

where $k_B$ and $T \sim 300K$ are the Boltzmann constant and the equilibrium temperature of a cell respectively. Thus, even the smallest living cells with the characteristic dimensions of $10^{-6} m$, are insensitive to thermal fluctuations at ambient conditions [23]. Remarkably, the value of $\frac{E_S}{k_B T} \cong \frac{\gamma_{int} l^2}{k_B T}$ may be identified with the information capacity of a single cell, according to the Landauer principle [24-25]. Landauer demonstrated that for creating (or erasing) of one bit of information at least $k_B T$ of energy should be spent (creating/erasing of information was treated by Landauer as a dissipative process, see Refs. 24-25). This idea was successfully verified experimentally in Ref. 26. If we speculate, that the information exchange occurs only via a surface of a living cell, Eq. 5b suggests an estimation of the informational capacity of a cell expressed in bits. The minimal value of this capacity for $l \cong 1 \mu m; \gamma_{int} \cong 1.0 \times 10^{-3} \frac{J}{m^2}$ equals $2.5 \times 10^5$ bits, as follows from Eq. 5b. However, for smaller cells, namely, when $l \cong 0.1 \mu m; \gamma_{int} \cong 1.0 \times 10^{-3} \frac{J}{m^2}$ we may estimate the informational capacity of a cell as $2.5 \times 10^3$ bits; in contrast DNA-based code enables the storage of $5.2 \times 10^6$ bits of information, as reported in Ref. 27. So it is reasonable to suggest, that the lower limit of the spatial dimensions of living cells is established not only by their stability in relating to thermal fluctuations, but also by their needs for informational exchange.

It is noteworthy, that the Landauer principle was recently criticized, and in particular in its applications to living cells and other biological systems [28-29]. So the estimation, supplied by Eq. 5b remains highly debatable. At the same time, the recent extended discussion of the applicability of the Landauer principle, presented in Ref. 30, demonstrated that the Landauer principle is generally true, and may be involved for rough estimations of the informational capacity of various systems.

*2.4. Spatial dimensions of living cells and kinetic considerations: interrelation between the temporal scales of mass and heat transport processes*

Additional arguments clarifying the characteristic spatial scales of living cells come from the analysis of mass transport taking place within a cell in its relation to chemical reactions occurring in cells. There exist at least two characteristic times appearing in the analysis of mass transport and the chemical events inherent for living cells. They arise from the analysis of the Brownian motion of particles (macromolecules or vesicles). The Brownian motion of a particle occurs under kinetics, described by the Fokker-Planck equation, characterized by the following time scales [9]:

$$\tau_{mix} \cong \frac{l^2}{D}, \tag{6a}$$

$$\tau_{traffic} \cong \frac{l^3}{DR}, \tag{6b}$$

where $\tau_{mix}$ and $\tau_{traffic}$ are the so-called "mixing" and "traffic" times, $D$ is the coefficient of diffusion, $R$ is the radius of the target, which a Brownian particle should come into contact with under random walking [9]. More accurately, $\tau_{mix}$ is the characteristic time of diffusion, whereas $\tau_{traffic}$ is the mean waiting time for a particular event which consists of touching a target with the characteristic dimension of $R$ by a particle walking randomly [9]. This event may be, for example, the migration of one enzyme to another. It is easily seen from Eqs. 6a-b, that the following interrelation takes place:

$$\frac{\tau_{traffic}}{\tau_{mixing}} \cong \frac{l}{R} \tag{7}$$

Quite expectably we conclude from Eq. 7, that $\tau_{traffic} \gg \tau_{mix}$ is valid. As shown in Ref. 9, for cells of micrometer size at room temperature $\tau_{traffic} \cong 1s$ is true; in larger cells (namely $l \cong 10\mu m$), the estimation supplies $\tau_{traffic} \cong 1000s$, when $R$ is taken as the characteristic dimension of a molecule with a medium molecular weight. Consider the characteristic times of enzymic reactions, in which the linking coenzyme has to migrate from one enzyme to another and back again in order to complete the catalytic cycle [31]. For a cell of a micrometric size this time is about one second. Thus, if the cell contains a single travelling-enzyme molecule and a single target-enzyme t molecule the rate of the reaction will be on the order of magnitude of one second. For a cell containing a few hundred target enzymes, the traffic time will be smaller by a factor of a few hundred, namely it will be on the order of magnitude of a millisecond, which coincides with the characteristic time scale of a single round of the catalytic enzyme cycle. This means that micron-scaled cells may be operated by a very small number of specific molecules, as it was observed in Ref. 9: "Small reaction subsystems of cell function *as coherent molecular networks* where individual reaction events, involving single molecules, are rigidly correlated and form causally related chains and loops." It is seen from Eq. 6b that $\tau_{traffic}$ scales as $l^3$, and for the cells with a size of $l \geq 100 \mu m$, the coherent control of cells by a small number of molecules will become impossible.

Until now we have considered only the time scales of mass transport. Now address the time scale of heat transport within a cell, supplied by:

$$\tau_{therm} \cong \frac{l^2}{\kappa}, \qquad (8)$$

where $\kappa \cong 1.2 \times 10^{-7} \frac{m^2}{s}$ and $D \cong 2.0 \times 10^{-5} \frac{m^2}{s}$ are the thermal diffusivity and the diffusion coefficient respectively, which are typical for living cells [32]. It is easily recognized, that following interrelation between $\tau_{mix}$ and $\tau_{therm}$ takes place:

$$\frac{\tau_{mix}}{\tau_{therm}} \cong \frac{\kappa}{D} \cong 6 \times 10^{-3}. \qquad (9)$$

Combining Eq. 7 and 9 generates the following hierarchy of time scales:

$$\tau_{traffic} >> \tau_{mix} >> \tau_{therm}. \qquad (10)$$

Equation 10 means that a thermal equilibrium is established immediately and heat transport does not impact the coherence of the diffusion-controlled processes (including chemical reactions) occurring in living cells.

*2.5. Spatial dimensions of living cells and characteristic lengths of instabilities occurring in liquid layers*

The narrow range of spatial dimensions of living cells $1 \mu m < l < 100 \mu m$ coincides with the range of dimensions of patterns arising from self-assembly processes observed in thin layers of polymer solutions [33-37]. A typical pattern formed under rapid evaporation of a polymer solution is depicted in Figure 1. The characteristic dimensions of cells in the pattern shown in Figure 1 are close to 20 μm. The processes of self-assembly taking place in liquid layers exposed to thermal gradients may be responsible for the genesis of living cells (although of course this hypothesis is purely speculative). The existence of giant living cells such as giant internodal cells (up to 5cm long), observed in *Characean Algae* calls for additional physico-chemical insights [38].

*2.6. Spatial dimensions of living cells and "the arrow of time"*

The origin of time irreversibility, called also "the arrow of time" remains highly debatable [39-44]. The arguments, supplied in Section 2.3 demonstrate that living cells should necessarily be macroscopic objects. Living cells, are not only macroscopic but also *dissipative* entities [45-46]. The macroscopic spatial dimensions of cells prevent them from destruction by thermal fluctuations and provide them with sufficient informational capacity, as established by the Landauer principle [24-26, 30]. It appears that the idea that living cells should necessarily be macroscopic entities (whose "large" spatial dimensions ensure their life-sustaining activities) was supposed by P. Teilhard de Chardin [47]. The fact, that living cells are dissipative structures [45-46, 48], built from a large number of molecules, guarantees the validity of the Second Law of Thermodynamics, giving rise to the "arrow of time" [23]. Therefore, we conclude that the phenomenon of "the arrow of time" occurs at the interface of physics and biology, due to the macroscopic spatial dimensions of living cells.

**Conclusions**

Spatial dimensions of typical living cells are confined within the narrow range, of between 1 and 100 μm. These dimensions arise from the conjunction of physical (thermodynamic and kinetic), chemical and biological reasonings. This means, that in living cells the volume (bulk) internal energy always prevails on over the surface energy, thus enabling the existence of living cells as physical entities. The upper limit of the spatial dimensions of living cells is restricted by gravity and inertia forces. The Bond number of millimetrically scaled cells is larger than unity. This means that effects due to gravity/inertia cannot be compensated by interfacial effects, and cells would necessarily be strongly deformed by gravity (or by acceleration), resulting in disruption of their functioning. Additional experimental efforts, devoted to establishment of the behavior of accelerated living cells are necessary (the experiments carried out with the centrifugal adhesion balance may shed a light on the problem [49]).

Kinetic arguments also restrict the upper limit of the dimensions of living cells. The analysis of the "traffic time" for the mass transport leads to the conclusion that coherent control of cells by a small number of molecules will be problematic for cells larger than 100 μm. For micron-scaled cells, the "traffic time" is on the order of magnitude of a millisecond, which coincides with the characteristic time scale of a single round of the catalytic enzyme cycle.

The lower limit of the spatial dimension of cells is defined by their ability to withstand thermal fluctuations on the one hand, and their need for a sufficient informational capacity on the other. Thus, we conclude that living cells are small, yet, essentially macroscopic, dissipative [45-46, 48] objects, built from a number of molecules, providing the applicability of the Second Law of Thermodynamics, giving rise to "the arrow of time".

**Acknowledgements**

The authors are indebted to Mrs. Ye. Bormashenko for her inestimable help in preparing this manuscript.

### References

1. Phillips, R.; Kondev, J.; Theriot, J. *Physical Biology of the Cell,* 2$^{nd}$ ed.; Garland Science, Taylor & Francis Group, LLC, London, UK, 2013.
2. Waigh, T. A. *The Physics of Living Processes: A Mesoscopic Approach*; Wiley, Chichester, UK, 2014.
3. Lim, C. T.; Zhou, E. H.; Quek, S. T. Mechanical models for living cells—a review. *J. Biomechanics* **2006**, *39 (2),* 195–216.
4. Wang, N.; Naruse, K.; , Stamenović, D.; Fredberg, J. J.; Mijailovich, S. M.; Tolić-Nørrelykke, I. M.; Polte, Th.; Mannix, R. Ingber, D. E. Mechanical behavior in living cells consistent with the tensegrity model. *PNAS¶* **2001**, *98 (14)*, 7765–7770.
5. Owicki, J. C.; Parce, J. W. Biosensors based on the energy metabolism of living cells: The physical chemistry and cell biology of extracellular acidification. *Biosensors & Bioelectronics*, **1992**, *7 (4)*, 255-272.

6. Qian, H.; Beard, D. A. Thermodynamics of stoichiometric biochemical networks in living systems far from equilibrium. *Biophysical Chemistry* **2005**, *114*, 213–220.

7. Davidson, R, M.; Lauritzen, A.; Seneff, St. Biological water dynamics and entropy: a biophysical origin of cancer and other diseases. *Entropy* **2013**, *15*, 3822-3876.

8. Davies, P. C. W.; Rieper, E.; Tuszynskic, J. A. Self-organization and entropy reduction in a living cell. *BioSystems* **2013**, *111*, 1– 10.

9. Hess, B.; Mikhailov. A. Self-organization in living cells. *Berichte der Bunsengesellschaft für physikalische Chemie* **1994**, *98 (9)*, 1198–1201.

10. Zhang, K.; Sejnowski, T. J. A Universal scaling law between gray matter and white matter of cerebral cortex. *PNAS* **2000**, *97 (10)*, 5621-5626.

11. Fabry, B.; Maksym, G. N.; Butler, J. P.; Glogauer, M.; Navajas, D.; Taback, N. A.; Millet, E. J.; Fredberg J. J. Time scale and other invariants of integrative mechanical behavior in living cells. *Phys. Rev. E* **2003**, *68*, 041914.

12. Gilbert, S. F. *Developmental Biology*; Sinauer Associates Inc. Massachusets, USA, 2000.

13. Marquet, P.; Rappaz, B.; Magistretti, P. J.; Cuche, T.; Emery, Y. Colomb, T.; Depeursinge, C. Digital holographic microscopy: a noninvasive contrast imaging technique allowing quantitative visualization of living cells with subwavelength axial accuracy. *Opt. Lett*. **2005**, *30 (5)*, 468-470.

14. Dumont, F.; Marechal, P.-A.; Gervais, P. Cell size and water permeability as determining factors for cell viability after freezing at different cooling rates. *Appl. Environ. Microbiol*. **2004**, *70*, 268-272.

15. McDonald, I. K.; Thornton, J. M. Satisfying Hydrogen bonding in proteins. *J. Mol. Biol.* **1994**, *238*, 777-793.

16. Foty, R.A.; Forgacs, G.; Pfleger, C. M.; Steinberg, M. S. Liquid properties of embryonic tissues: measurement of interfacial tensions. *Phys. Rev. Lett*. **1994**, *72 (14)*, 2298-2301.

17. Morris, C. E.; Homann, U. Cell surface area regulation and membrane tension. *J. Membrane Biol.* **2001**, *179*, 79–102.

18. de Gennes, P. G.; Brochard-Wyart, F.; Quéré, D. *Capillarity and Wetting Phenomena;* Springer, Berlin, 2003.

19. Erbil, H. Y. *Surface Chemistry of Solid and Liquid Interfaces*, Blackwell, Oxford, 2006.
20. Bormashenko, Ed. *Wetting of Real Surfaces, de Gruyter*, Berlin, 2013.
21. He, L.; Luo, Z. Y.; Xu, F.; Bai, B. F. Effect of flow acceleration an deformation and adhesion dynamics of captured cells. *J. Mechanics in Medicine and Biol*. **2013**, *13(5)*, 1340002.
22. Cheung, L. S. L.; Zheng, X.; Stopa, A.; Baygents, J. C.; Guzman, R.; Schroeder, J. A.; Heimark R. L.; Zohar, Y. Detachment of captured cancer cells under flow acceleration in a bio-functionalized microchannel. *Lab on Chip* **2009** *9 (12)*, 1721-1731.
23. Landau, L.; Lifshitz, E. M. *Statistical Physics*, vol. 5 (3rd ed.), Butterworth-Heinemann, Oxford, UK, 1980.
24. Landauer, R. Irreversibility and heat generation in the computing process. *IBM J. Res. Develop.* **1961**, *5*, 183-191.
25. Landauer, R. Dissipation and noise immunity in computation and communication. *Nature* **1988**, *355*, 779-784.
26. Berut, A.; Arakelyan, A.; Petrosyan A.; Ciliberto, S.; Dillinschneider, R.; Luts, E. Experimental verification of Landauers principle linking information and thermodynamics. *Nature*, **2012**, *483*, 187-189.
27. Goldman, N.; Bertone, P.; Chen, A.; Dessimoz, C.; LeProust, E. M.; Sipos, B.; Birney, B. Towards practical, high-capacity, low-maintenance information storage in synthesized DNA. *Nature*, **2013**, *494* , 77–80.
28. Li, J. Planckian Information ($I_p$): a new measure of order: in atoms, enzymes, cells, brains, human societies, and Cosmos, in *Unified Field Mechanics: Natural Science beyond the Veil of Spacetime* (R. Amoroso, P. Rowlands, and L. Kauffman, L. eds.), World Scientific, New Jersey, USA, 2015, pp. 579-589.
29. Li, J. Planckian distributions in molecular machines, living cells, and brains : The wave-particle duality in biomedical sciences. *New Developments in Biology, Biomedical & Chemical Engineering and Materials Science*, Proceedings of the International Conference on Biology and Biomedical Engineering (BBE 2015), ed. by: M. Razeghi, V. Mladenov, P. A. Anninos), Vienna, Austria, March, 15-17, **2015**, pp. 115-137.
30. Lutz, E.; Ciliberto, S. Information: from Maxwell's demon, to Landauer eraser, Physics Today, 2015, 68, 30-35.

31. Dixon, M.; Webb, E. C.; Thorne, C. J. R.; Tipton, R. F. *Enzymes*, 3rd Ed. Longmans Group Ltd., London, UK, 1979.

32. Svaasand, L. O.; Boerslid, T.; Oeveraasen, M. Thermal and optical properties of living tissue: Application to laser-induced hyperthermia. *Lasers in Surgery & Medicine*, **1985**, *5 (6)*, 589–602.

33. Koschmieder, E. L. *Bénard Cells and Taylor Vortices*, Cambridge University Press, Cambridge, UK, 1993.

34. Nepomnyashchy, A. A.; Velarde, M. G., Colinet, P. *Interfacial phenomena and convection,* CRC Press, Boca Raton, USA, 2001.

35. de Gennes, P. G.; Brochard-Wyart, F.; Quéré, D. *Capillarity and Wetting Phenomena*, Springer, Berlin, 2003.

36. Bormashenko, Ed.; Pogreb, R.; Musin, A.; Stanevsky, O.; Bormashenko, Ye.; Whyman, G.; Gendelman, O.; Barkay, B. Self-assembly in evaporated polymer solutions: Influence of the solution concentration. *J. Colloid & Interface Sci*. **2006**, *297 (2)*, 534-540.

37. Bormashenko, Ed.; Bormashenko, Ye.; Gendelman, O. On the nature of the friction between nonstick droplets and solid substrates. *Langmuir* **2010**, *26 (15)*, 12479-12482.

38. Braun, M.; Foissner, I.; Löhring, H.; Schubert, H.; Thiel, G. Characean Algae: Still a Valid Model System to Examine Fundamental Principles in Plants. Chapter, *Progress in Botany*, Vol. 68, of the series Progress in Botany, Springer, pp. 193-220, **2007**, Doi: 10.1007/978-3-540-36832-8_9

39. Halliwell, J. J.; Perez-Mercader, J.; Zurek, J. H. *Physical Origins of Time Asymmetry*, Cambridge University Press, Cambridge, UK. 1994.

40. Maccone, L. Quantum solution to the arrow-of-time dilemma, *Phys. Rev. Lett*. **2009**, 103, 080401.

41. Mikhailovsky, F. E.; Levich, A. P. Information and complexity or which aims the arrow of time? *Entropy* **2015**, *17(7)*, 4863-4890;

42. Aharonov, Y.; Cohen, T.; Landsberger, T. The Two-time interpretation and macroscopic time-reversibility. *Entropy* **2017**, *19(3)*, 111; doi:10.3390/e19030111

43. Moffat, J. W. Gravitational entropy and the Second Law of Thermodynamics *Entropy* **2015**, *17(12)*, 8341-8345.

44. Nosonovsky, M.; Mortazavi H. *Friction-Induced Vibrations and Self-Organization: Mechanics and Non-Equilibrium Thermodynamics of Sliding Contact*, CRC Press, Fl. Boca Raton, USA, 2014.

45. Petty, H, R.; Worth, R. G.; Kindzelskii, A. L. Imaging Sustained Dissipative patterns in the metabolism of individual living cells. *Phys. Rev. Lett*. **2000**, *84*, 2754.

46. Lim, C. T.; Zhou, E. H.; Quek, S. T. Mechanical models for living cells—a review. *J. Biomechanics*, **2006**, *39 (2)*, 195–216.

47. Teilhard de Chardin, P. *Le Phénomène humain (in French)*, Editions du Seuil, Paris, France, 1956.

48. Wang, J.; Xu, Li.; Wang, Er.; Huang, S. The Potential landscape of genetic circuits imposes the arrow of time in stem cell differentiation. *Biophysics J*. **2010**, *99 (1)*, 29–39.

49. Tadmor, R.; Bahadur, Pr.; Leh, A.; N'guessan, H. T.; Jaini, R.; Dang, L. Measurement of lateral adhesion forces at the interface between a liquid drop and a substrate. *Phys. Rev. Lett*. **2009**, *103*, 266101.

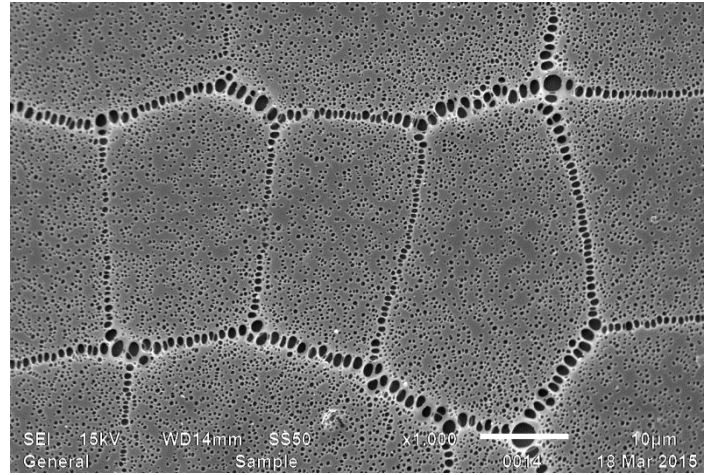

Figure 1. Typical pattern observed under rapid evaporation of polymer solutions [33-34]. Scale bar is 10 μm.